\documentclass[letter]{aa} 

\usepackage{graphicx}
\usepackage{txfonts}
\begin{document}

   \title{Study of a new central compact object: The neutron star in 
     the supernova remnant G15.9+0.2}


   \author{D.~Klochkov\inst{1} \and
          V.~Suleimanov\inst{1}\and
          M.~Sasaki\inst{1} \and
          A.~Santangelo\inst{1}
          }

   \institute{Institut f\"ur Astronomie und Astrophysik (IAAT), Universit\"at
     T\"ubingen, Sand 1, 72076 T\"ubingen, Germany
   }

   \date{Received ***; ****}

 
  \abstract
  {We present our study of the central point source
    CXOU\,J181852.0$-$150213 in the young Galactic
    supernova remnant (SNR) G15.9+0.2 based on the recent $\sim$90\,ks 
    \emph{Chandra} observations. The point source was
    discovered in 2005 in shorter \emph{Chandra} observations and was
    hypothesized 
    to be a neutron star associated
    with the SNR. Our X-ray spectral analysis strongly supports 
    the hypothesis of a thermally emitting neutron star associated
    with G15.9+0.2. We conclude that the object belongs to the class
    of young cooling low-magnetized neutron stars referred to as
    central compact objects (CCOs). We modeled the spectrum of the
    neutron star with a blackbody spectral function and with our
    hydrogen and carbon neutron star atmosphere models, assuming
    that the radiation is uniformly emitted by the entire stellar
    surface. Under this assumption, only the carbon atmosphere models
    yield a distance that is compatible with a source located in the Galaxy.
    In this respect, CXOU\,J181852.0$-$150213 is similar to two other
    well-studied CCOs, the neutron stars in Cas~A and in
    HESS\,J1731$-$347, for which 
    carbon atmosphere models were used to reconcile their emission
    with the known or estimated distances.
  }


   \maketitle
%

\section{Introduction}

Neutron stars are fundamental objects of modern astrophysics
because (i) supranuclear densities in their interiors
probably yield unexplored forms of matter, (ii) they are the outcome of 
the explosive collapse of massive stars, and (iii) their extreme gravitational
and magnetic fields give rise to reach phenomenology of the associated
emission and accretion. Only a few dozen neutron stars (NSs) are associated
with their parent supernova remnants (SNRs). A dozen of these stars show pure
thermal X-ray emission 
and lack any magnetic activity such as flaring, nonthermal
magnetospheric emission or pulsar wind nebulae. This subgroup are
usually referred to as central compact objects (CCOs) and were first
considered as a separate class of isolated NSs by
\citet{Pavlov:etal:02,Pavlov:etal:04} based on \emph{Chandra}
observations. For three CCOs, pulsations with periods in the range of
$\sim$0.1--0.4\,s have been detected. 
The measured spin-down rates
 confirm relatively low magnetic
fields of these objects, namely,
$B\sim 3\times 10^{10}- 10^{11}$\,G
\citep{Halpern:Gotthelf:10,Gotthelf:etal:13},
which is substantially weaker than in radio and accreting pulsars.
The class of CCOs is thus believed to be composed by young
(a few 10$^3-10^4$\,yr), weakly magnetized, thermally emitting,
cooling NSs. 

One of the major efforts in NS studies is the modeling of the
thermal emission from the stellar surface. Under certain assumptions,
such a modeling permits constraints on
the geometrical and physical properties of the stars and, most importantly, on their mass, radius, and effective temperature. Constraints on the
mass and radius allow one to probe the equation of state of the
superdense matter in the stellar interior 
\citep[e.g.,][]{Haensel:etal:07,Lattimer:Prakash:15,Watts:etal:16}.
The effective temperature of the stellar surface combined with the
estimated age of the object allows one to probe the NS
cooling rate \citep[e.g.,][]{Page:etal:04,Page:etal:06}. 
The cooling mechanisms also strongly depend on the
microphysics inside the star
\citep[e.g.,][]{Ofengeim:etal:15,Beznogov:Yakovlev:15}.   
The relatively weak magnetic fields of CCOs permit modeling of
their thermal emission under simplified assumptions, namely neglecting
the effects of strong magnetic fields in the stellar atmosphere
\citep[e.g.,][and references therein]{Suleimanov:etal:14}; this makes
these objects perfect laboratories for the study of neutron star physics.

The point source  \object{CXOU\,J181852.0$-$150213} (hereafter,
CXOU\,J1818) was serendipitously discovered in the \emph{Chandra}
X-ray observations 
of the radio-bright, few thousand year old SNR \object{G15.9+0.2}
\citep{Reynolds:etal:06}. The relatively sparse data did not permit a
detailed spectral analysis of the source. \citet{Reynolds:etal:06}
argued, however, that CXOU\,J1818 is probably a NS associated with the
remnant. A rotation-powered pulsar or a CCO were considered by the
authors to be viable
possibilities. The object was subsequently listed among ``CCO
candidates'' in \citet{Gotthelf:etal:13}. In this Letter, we present
our analysis of the new \emph{Chandra} observation of G15.9+0.2 and
CXOU\,J1818, which has increased 
the total exposure time for these sources by a factor of four. Our
study is focused on the compact source and is fully consistent with the CCO
hypothesis. The detailed analysis of the diffuse emission of the SNR
is not part of this work and will be presented elsewhere.

\section{Observations and data analysis}

\begin{figure}
\centering
\resizebox{\hsize}{!}{\includegraphics{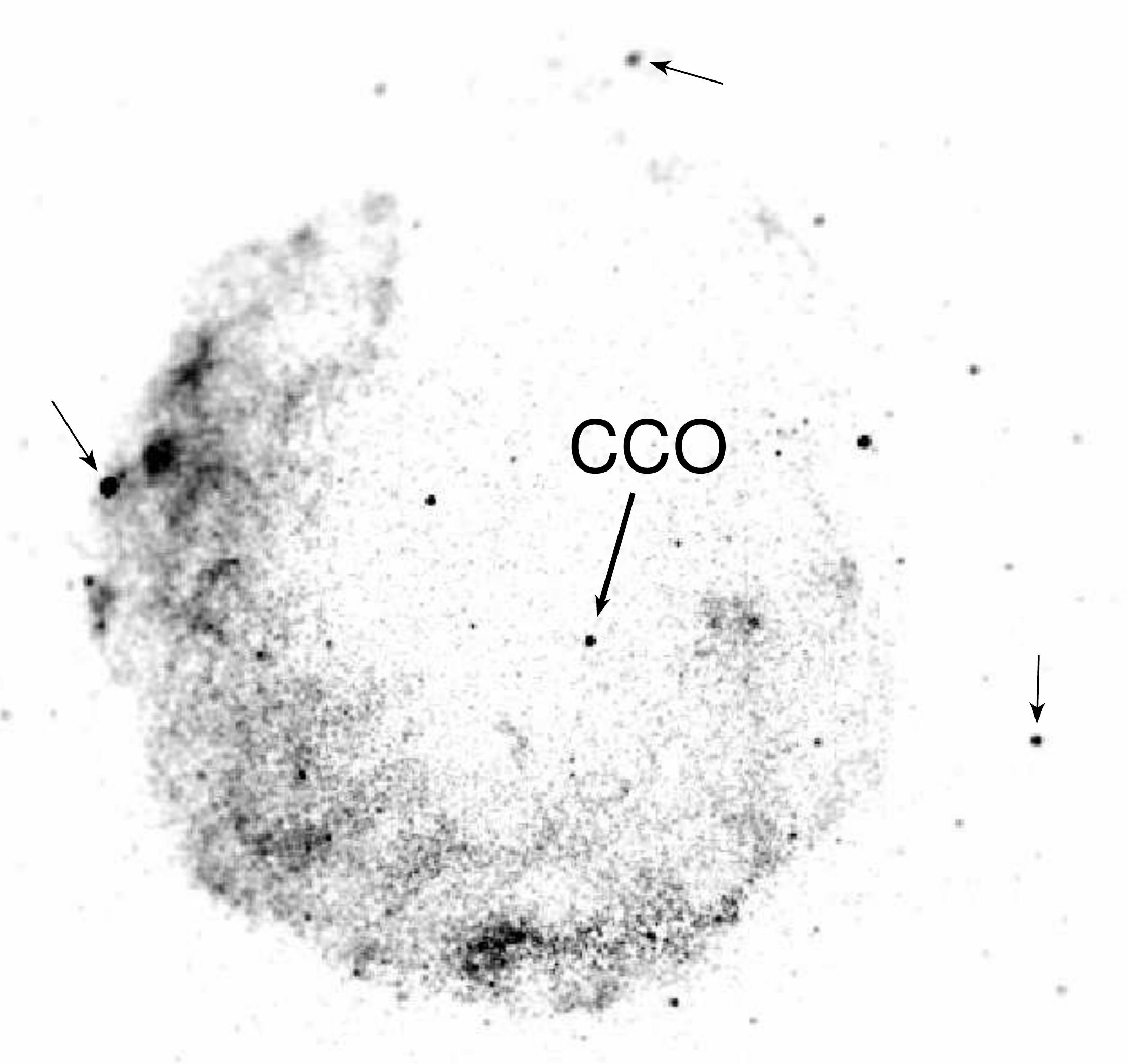}}
\caption{Background-subtracted, exposure-corrected ACIS image of
  G15.9+0.2 in the 0.5--7\,keV range extracted from our 2015 data. The
  central source CXOU\,J181852.0$-$150213 is indicated as ``CCO''. The narrow
  arrows indicate the three foreground stars used to correct the
  coordinate system in the image.}  
\label{fig:ima}
\end{figure}

The supernova remnant G15.9+0.2 was observed with 
the ACIS detector on board \emph{Chandra} 
on June 30-31, 2015 for 93\,ks (Obs. ID 16766). For the data
processing and analysis, we used
version 4.8 of Chandra Interactive Analysis of
Observations software (CIAO; \citealt{Fruscione:etal:06}) with the
corresponding CALDB version 4.7.0. We applied the optional VFAINT
correction for a more effective elimination of background events.
A light curve extracted from a source-free region was used to
check for possible background 
flares\footnote{\url{http://cxc.harvard.edu/contrib/maxim/bg/index.html}};
no flares were identified in our observations.
Most of the remnant is covered by the back-illuminated S3 CCD-chip.
Figure\,\ref{fig:ima} shows a background-subtracted
exposure-corrected image of G15.9+0.2 in the 0.5--7\,keV range. 
We used the background from a set of \emph{blank-sky} observations
provided as a part of \emph{Chandra} CALDB. The final image was
smoothed with a local PSF at each point. 

The point source CXOU\,J1818 is clearly visible in the ACIS image. 
For the spectral extraction, we used a
circle with a radius of 1.2 arcsec encompassing $\sim$90\% of the source photon
energy. In total, 307 photons are detected within the extraction radius.
The background spectrum is extracted from an annulus around the source
position with an inner and an outer radii of 2 and 10 arcsec, respectively. 
The exposure time of single frames was 3.14\,s, which translates into
$\sim$0.01 source counts per frame in our observations. For such a count
rate, the expected pile-up fraction is expected to be below
one percent\footnote{\url{http://cxc.harvard.edu/csc/memos/files/Davis_pileup.pdf}}.

We also reanalyzed the data of CXOU\,J1818 from the 2005
\emph{Chandra} observations ($\sim$30\,ks) presented by
\citet{Reynolds:etal:06}. The obtained spectral parameters are found
to be consistent 
with those presented by the authors as well as with those obtained
with the new observations. Therefore, for our final results, we used
simultaneous fits of the spectra extracted with the old and new
observations. 

\section{Absolute astrometry and timing}

With the new observations, we attempted an improvement of the sky
coordinates of CXOU\,J1818. The coordinates measured with the CIAO tool
\texttt{wavedetect} in our observations 
have uncertainties of $\sim$0.02\,arcsec.
The systematic uncertainty of the derived sky positions related to 
inaccuracy of the determined instrument pointing direction
is expected to be 0.4 arcsec according to the instrument
team\footnote{\url{http://cxc.harvard.edu/ciao/threads/reproject_aspect/}}.
We therefore performed a correction of the measured coordinates for a possible
systematic offset. We  identified three stars from the fourth
U.S. Naval Observatory CCD Astrograph Catalog (UCAC4), which are visible in our
\emph{Chandra} image (indicated with thin arrows in
Fig.\,\ref{fig:ima}). Our correction of the instrument pointing 
direction based on the optical sky positions of the stars resulted in
a total shift of the coordinate system in the ACIS frame by 
0.23\,arcsec. The mean residual offset between the optical and X-ray 
positions of the three stars after the correction turns out to be
0.15\,arcsec, which is within the statistical uncertainties of the measured
X-ray coordinates of the stars. We thus conservatively ascribe 
a total uncertainty of 0.2\,arcsec to the corrected sky
coordinates of CXOU\,J1818 (J2000): 
$\alpha=18$h\,18m\,52.080s, 
$\delta=-15^\circ$\,02$^\prime$\,14.11$^{\prime\prime}$. 

We compared the updated coordinates of the CXOU\,J1818 with those of the
two nearby stars from the \emph{Spitzer's} infrared GLIMPSE survey,
G015.8781+00.1969 and G015.8774+00.1965, which are considered by
\citet{Reynolds:etal:06} as possible counterparts of the X-ray source. The
angular separations with G015.8781+00.1969 and G015.8774+00.1965 are 
2.1 and 0.8\,arcsec, respectively. 
Comparing these separations with the uncertainty of 0.2~arcsec derived above
we conclude, similarly to \citet{Reynolds:etal:06}, that the two stars
are unlikely counterparts of CXOU\,J1818. 

The low photon statistics ($\sim$300 source photons) does not permit a
detailed timing analysis of the source. Nevertheless, we
performed a search for pulsations down to the 
pulse period of $\sim$6\,s corresponding to the Nyquist frequency
determined by the time resolution of the data ($\sim$3\,sec) using 
the Rayleigh periodogram \citep[$Z_1^2$-statistics; e.g.,][and references
therein]{Protheroe:etal:87}. The derived 99\% c.l. upper limit on the amplitude
of sinusoidal pulsations turns out to be 0.56. Since the background is
negligible, the derived upper limit essentially applies to the
intrinsic pulsed fraction of the source.

\section{Spectral modeling}

The new observation of CXOU\,J1818 has a factor of three
longer exposure time compared to the 2006
observations. The combined spectrum permits a detailed
comparison between the spectral models of
absorbed blackbody and absorbed power law. 
Such a comparison is crucial to establish the thermally
emitting NS nature of the source. We rebinned the spectrum
allowing a minimum of 20 photons per bin, 
which permits the usage of the Gaussian statistic
  ($\chi^2$) in the spectral modeling. The adopted number of counts per
  bin is, however, at the verge of the $\chi^2$ applicability. We have,
therefore, redone all our spectral fits using the Poisson statistic
(\texttt{cstat} in XSPEC, \citealt{Cash:79}). The differences in the
best-fit parameters between the two statistics are negligible compared
to the statistical uncertainties of the parameters. All results of
the spectral fits reported below are obtained with the $\chi^2$ statistic.
To model the interstellar absorption at lower energies, we used the
XSPEC \texttt{wabs} model with Wisconsin cross-sections 
\citep{Morrison:McCammon:83} and the Anders \& Ebihara relative
abundances \citep{Anders:Ebihara:82}.

\begin{figure}
\centering
\resizebox{\hsize}{!}{\includegraphics{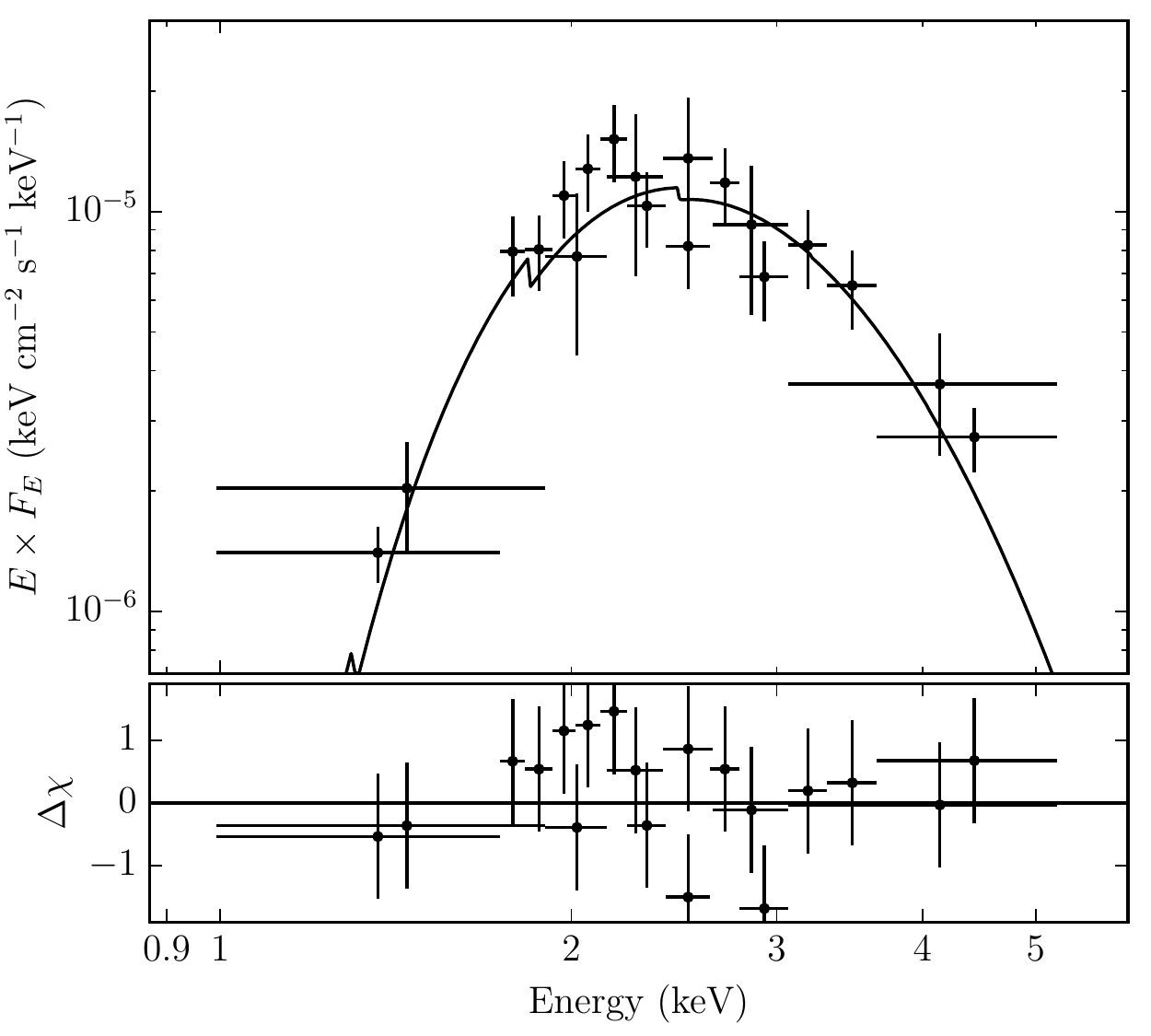}}
\caption{Unfolded \emph{Chandra}/ACIS spectra of the central point source
  CXOU\,J181852.0$-$150213
  obtained in the 2005 and 2015 observations fitted with an absorbed
  blackbody model.
}
\label{fig:spe}
\end{figure}

Both models, absorbed blackbody and absorbed power law,
provide formally acceptable fits to
the data, with $\chi^2_{\rm red}\simeq 0.9$ and 0.7,
  respectively, for 15 degrees of 
  freedom (d.o.f.). The power-law model results, however, in a photon index of
$\Gamma\sim 6,$ which is substantially higher than the
photon indices typical for the spectra of AGNs, 
young radio pulsars, and of pulsar wind nebulae. A blackbody fit on the
other hand yields a temperature of $kT\sim 0.5$\,keV
(Fig.\,\ref{fig:spe}), which is
typical for central neutron stars in SNRs \citep[e.g.,][]{Pavlov:etal:04}. 
The absorbed flux in the 1$-$5\,keV energy range (no photons are observed
outside this range) is $(3.6\pm 0.2)\times 10^{-14}$\,erg~s$^{-1}$~cm$^{-2}$.

Different spectral functions chosen to fit the data of CXOU\,J1818,
namely power law versus blackbody, lead to different
constraints on the absorption column density $n_{\rm H}$. This value can be
compared with the $n_{\rm H}$ value measured for the diffuse SNR emission,
which has much better photon statistics compared to the point source. 
We modeled the entire SNR spectrum using the nonequilibrium
ionization collisional plasma model \texttt{vnei} of XSPEC. 
We had to introduce an artificial nonzero redshift $z\simeq 8\times
10^{-3}$, which apparently accounts for an uncalibrated gain drift of
the ACIS energy
scale\footnote{\url{http://cxc.harvard.edu/cal/Acis/Cal_prods/tgain/index.html}}. 
As a result, an absorption column density of 
$n_{\rm H}=(3.5\pm 0.1)\times 10^{22}$\,cm$^{-2}$ 
is measured. We  also extracted 
SNR spectra from a number of different regions of the remnant. The
measured absorption varies in a range $(3.2-4.4)\times
10^{22}$\,cm$^{-2}$, whereas a value of 
$(3.3\pm 0.1)\times 10^{22}$\,cm$^{-2}$ is measured at the central region of
the remnant surrounding CXOU\,J1818. A blackbody fit to
the point source spectrum 
yields $3.4_{-0.4}^{+0.5}\times 10^{22}$\,cm$^{-2}$ , which is fully consistent with the
SNR values. The power-law fit to CXOU\,J1818, however, results in
$n_{\rm H}=(6.6\pm 0.7)\times 10^{22}$\,cm$^{-2}$ ; this is significantly higher
than the absorption of the diffuse emission. It is therefore unlikely that the
central source has an intrinsic power-law spectrum. Its spectral properties are
much more typical for a thermally emitting NS as in the case of
CCOs.

\section{Neutron star atmosphere models}

The blackbody fit to the spectrum of
CXOU\,J1818 leads to an unrealistic distance to the object, which is on the
order of 200\,kpc if we assume the radius of the emitting region to
be equal to the canonical NS radius of 12\,km. A realistic distance of
$\sim$10\,kpc or below would require a radius of the emitting region
below $\sim$0.5\,km. Such an apparent discrepancy
between the radii of the emitting regions derived from a blackbody
spectral model and, assuming the canonical radius of the NS, is known
for CCOs; see, e.g., Table~2 in \citet{Pavlov:etal:04}. A solution to
this problem is provided by calculations of the emergent spectrum
using NS atmosphere models 
that are sometimes combined with the assumption of compact
emitting ``hot spots'' on the stellar surface
\citep{Pavlov:Luna:09,Ho:Heinke:09,Klochkov:etal:13,Bogdanov:14}. 
We applied the nonmagnetic hydrogen and carbon atmosphere models
developed in our group for NSs
\citep{Suleimanov:etal:14,Klochkov:etal:15} to fit the
extracted ACIS spectrum of CXOU\,J1818\footnote{The carbon atmosphere model
  for XSPEC is available at
  \url{https://heasarc.gsfc.nasa.gov/xanadu/xspec/models/carbatm.html}}.  

\begin{table*}
  \centering
  \renewcommand{\arraystretch}{1.3}
  \caption{Results of the spectral modeling of
    CXOU\,J181852.0$-$150213 with a blackbody spectral function and
    with hydrogen and carbon atmosphere models. 
    The fits with the atmosphere models
    are provided for the two cases: with the absorption column
    density $n_{\rm H}$ as a free fit parameter and with $n_{\rm H}$
    fixed to the value of $3.5\times 10^{22}$\,cm$^{-2}$
    measured for the diffuse SNR emission. The mass and
    radius of the neutron star in the model atmosphere fits are fixed
    to to 1.5$M_\odot$ and 12\,km,
    respectively. The distances are derived from the normalization
    constant of the models assuming a stellar radius of 12\,km.
    The indicated uncertainties are at 1\,$\sigma$ c.l.}
  \label{tab:spe}
  \begin{tabular}{l|l|l|l|l|l}
    \hline\hline
  Parameter         & Blackbody  & \multicolumn{2}{c|}{Hydrogen} & \multicolumn{2}{|c}{Carbon}  \\
    \hline
                    &            &   & &       & \\
$n_{\rm H}/(10^{22}$\,atom~cm$^{-2})$ & $3.4_{-0.4}^{+0.5}$ & $4.2_{-0.5}^{+0.6}$ & 3.5 (fixed) & $4.3\pm 0.5$ & 3.5 (fixed)\\
 $T$, MK                &$5.5\pm 0.4$& $3.1\pm 0.4$ &$3.6\pm 0.2$& $1.9\pm 0.3$ &$2.4\pm 0.2$\\
Distance, kpc  & $216_{-54}^{+67}$ & $40_{-14}^{+21}$ & $66_{-8}^{+9}$ & $10_{-5}^{+9}$ & $26\pm 5$ \\
$\chi^2_{\rm red}$/d.o.f.  & 0.9/15 & 0.8/15 & 0.8/16 & 0.8/15 & 0.9/16 \\
    \hline
  \end{tabular}
\end{table*}

In our model atmosphere fits, we assume that the object is a NS
associated with the remnant. We, therefore,
fixed the absorption column density to the value of $3.5\times
10^{22}$\,cm$^{-2}$ obtained for the
diffuse emission of G15.9+0.2. The photon statistics in the spectrum 
are insufficient to obtain any meaningful
constraints on the mass and radius of the star. Therefore, we
fixed these parameters to the canonical values of $M=1.5M_\odot$ and
$R=12$\,km. The results of the fit with the atmosphere models and with
a blackbody spectral function are summarized in 
Table\,\ref{tab:spe}. A known effect of applying the atmosphere models
that can be recognized 
is a lower effective temperature compared to that obtained from the
blackbody fit \citep[e.g.,][]{Suleimanov:etal:14}. As a result, the
distances derived from the normalization constant of the models are
substantially reduced. For the carbon atmosphere model, it is consistent
with the source being located in the Galaxy (Table\,\ref{tab:spe}).

However, the value of the absorption column
  density derived from the modeling of the SNR diffuse emission might
  be affected by 
  additional systematic uncertainties related to the choice of the
  plasma emission model (we used the simplest
  single-temperature nonequilibrium plasma),
  an inaccurate background-subtraction, and to the inhomogeneity of the
  absorbing matter across the remnant. Therefore, in
  Table\,\ref{tab:spe} we also quote the results of the model
  atmosphere fits leaving $n_{\rm H}$ as a free fit parameter. The
  uncertainties of the spectral parameters, especially of the derived
  distances, which naturally increase in this case. Nevertheless, the
  mentioned conclusions of the model atmosphere fits are still valid; only
the carbon atmosphere model yields a distance that is formally compatible with a
Galactic SNR. The best-fit absorption column densities stay within the
range measured over the remnant (see the previous section).

The distances derived from the model atmosphere fits scale
  with the assumed mass and radius of the NS, which are fixed to the
  canonical values in the fits above. In the case of the carbon
  atmosphere model and fixed $n_{\rm H}$, the scaling is such that the best-fit
  distance increases to $\sim$30\,kpc for $M=2\,M_\odot$ and decreases
  to $\sim$20\,kpc for  $M=1\,M_\odot$. The scaling with the NS radius
  is such that the best-fit distance increases to $\sim$30\,kpc for $R=15$\,km
  and decreases to $\sim$22\,kpc for $R=10$\,km. For the hydrogen
  atmospheres, the scaling is in the same direction.


\section{Conclusions}

Our spectral analysis of the central source in the SNR G15.9+0.2
clearly supports the hypothesis of a thermally emitting NS. The object
thus can be considered to be
a new member of the CCO class. A power-law spectral model yields an
absorption column density that is incompatible with that of the remnant and a
photon index that is much softer than expected in case of an AGN, a young
radio pulsar, or a pulsar wind nebula. We thus consider these
alternatives as rather improbable. We also confirmed the absence of an
optical or infrared counterpart with the improved
sky coordinates of the source. The stellar coronal origin of the
object is, therefore, also unlikely. 

Similar to other CCOs, the distance to CXOU\,J1818 derived from the
blackbody normalization is unrealistically high, $\sim$200\,kpc,
assuming that the radiation is emitted by the entire stellar
surface. A fit with the hydrogen atmosphere models reduces the
distance to $\sim$40--70\,kpc, which is still too high for a
Galactic source. Only the carbon atmosphere model yields a
meaningful distance of $\sim$10--20\,kpc. Very similar
results have been 
obtained for two other CCOs for which much better observational data
are available (higher flux, longer and multiple observations), specifically, the
central NSs in SNRs Cas~A \citep{Ho:Heinke:09} and
HESS\,J1731$-$347 \citep{Klochkov:etal:13,Klochkov:etal:15}. In these
two objects, the 
hypothesis that the radiation is uniformly emitted by the entire
stellar surface (i.e., that the emitting radius is equal to the
stellar radius) is supported by the relatively stringent limits on the
pulsed fraction of $\lesssim$10\%. Our data do not permit any
useful constraints on pulsations of CXOU\,J1818. The pulse periods
of the known pulsations CCOs, RX\,J0822.0$-$4300, 1E\,1207.4$-$5209,
and CXOU\,J185238.6$+$004020, are all below $\sim$0.4\,s, which is in the
range inaccessible with the \emph{Chandra} observations due
to insufficient timing resolution. 
Nevertheless, the similarities in
the spectral properties of CXOU\,J1818 with the well-studied
CCOs in Cas~A and in HESS\,J1731$-$347 suggest 
a similar geometry and physics of emission in the three NSs
and further supports the association of CXOU\,J1818 with the
class of CCOs.

\begin{acknowledgements}
We thank the anonymous referee for her/his comments and suggestions
that substantially improved the manuscript.
M.S. acknowledges support by the German Research Foundation
through the Heisenberg grant SA 2131/3-1. 
V.S. acknowledges support by the German Research Foundation
through the grant WE 1312/48-1 
\end{acknowledgements}

\bibliographystyle{aa}
\bibliography{refs}

\end{document}